# DATA SMOOTHING FILLING METHOD BASED ON scRNA-SEQ DATA ZERO-VALUE IDENTIFICATION

Linfeng Jiang[1,2,3] and Yuan Zhu[1,2,3*]

[1]School of Automation, China University of Geosciences, Hongshan District, No. 388 Lumo Road, 430074, Wuhan, China
[2]Hubei Key Laboratory of Advanced Control and Intelligent Automation for Complex Systems, Hongshan District, No. 388 Lumo Road, 430074, Wuhan, China
[3]Engineering Research Center of Intelligent Technology for Geo-Exploration, Hongshan District, No. 388 Lumo Road, 430074, Wuhan, China
jianglinfeng@cug.edu.cn
zhuyuan@cug.edu.cn

## ABSTRACT

*Single-cell RNA sequencing (scRNA-seq) determines RNA expression at single-cell resolution. It provides a powerful tool for studying immunity, regulation, and other life activities of cells. However, due to the limitations of the sequencing technique, the scRNA-seq data are represented with sparsity, which contains missing gene values, i.e., zero values, called dropout. Therefore, it is necessary to impute missing values before analyzing scRNA-seq data. However, existing imputation computation methods often only focus on the identification of technical zeros or imputing all zeros based on cell similarity. This study proposes a new method (SFAG) to reconstruct the gene expression relationship matrix by using graph regularization technology to preserve the high-dimensional manifold information of the data, and to mine the relationship between genes and cells in the data, and then uses a method of averaging the clustering results to fill in the identified technical zeros. Experimental results show that SFAG can help improve downstream analysis and reconstruct cell trajectory.*

## 1. INTRODUCTION

Single-cell RNA sequencing (scRNA-seq) technology is a developing technology for amplifying and sequencing the whole transcriptome at a single cell level, revealing the heterogeneity of individual cells in a seemingly uniform cell population or tissue. It enables in-depth exploration of the characteristics and fate of cells[1,2]. For example, scRNA-seq can discover new subpopulations and expound the differentiation pathways of cells, revealing the relationship between intratumor transcriptome heterogeneity and predicted results[3,4]. However, scRNA-seq is susceptible to technical noise compared with bulk RNA-seq. This is called the "dropout" problem. A high dropout ratio may disrupt latent biological signals and affect downstream analyses such as low-latitude expression and clustering[5-7]. Therefore, it is necessary to design appropriate imputation methods for scRNA-seq data.

Recently, researchers have proposed many methods to impute the dropout events of scRNA-seq data from different perspectives[8]. These methods usually rely on similar cells or genes containing valuable information to predict gene expression. For example, MAGIC constructed the Markov transition matrix through cell similarity and estimated gene expression through the rich structure provided by biology. However, MAGIC disturbs gene distribution, resulting in over-smoothing the data [9]. scImpute used a mixed model to predict the probability of gene miss and used information from similar cells to estimate gene expression with a high miss ratio[10]. However, it imputed dropout events according to the clustering result, so they need to know the true value of the class in advance. SAVER used the relationship between genes to recover gene expression, which first performed a linear regression on the genes and then used the values calculated by the posterior distribution to impute. However, it costs much more time to calculate many parameters[11]. CMF-Impute used non-negative matrix factorization to impute the dropout events and integrated the similarity matrix of cells and genes into the objective function, thus CMF-Impute combined cell and gene information[12]. Identifying technical zeros is essential for imputing scRNA-seq data. It is equally important to use information from similar cells to fill in the technical zeros identified. Existing methods rarely take both into account.

In this study, we developed an imputation method (SFAG) for scRNA-seq data that simultaneously considers technical zeros and cell similarity. The method first integrates the similarity of cell and gene gene expression relationships as graph regularisation terms into a non-negative matrix decomposition, which is solved iteratively to discriminate between zero values in the data[13]; then, similar cells are identified by clustering and the expression values of similar cells are averaged to perform interpolation[14].

The experiments are conducted on two real datasets, and the results are compared with other eight state-of-the-art methods, including DrImpute[14], MAGIC[9], CNMF[12], ALRA[15], scRMD[16], G2S3[17], VIPER[18], SAVER[11]. For the real datasets, we used tSNE + K-means to cluster and calculated the evaluation index to evaluate the performance of each method[19].

## 2. MATERIALS AND METHODS

### 2.1. Overview: SFAG

This study proposes a novel method SFAG for imputing scRNA-seq data. As is shown in Figure 1, which includes Graph regularized non-negative matrix factorization, average clustering results to impute the data, and updating imputation results.

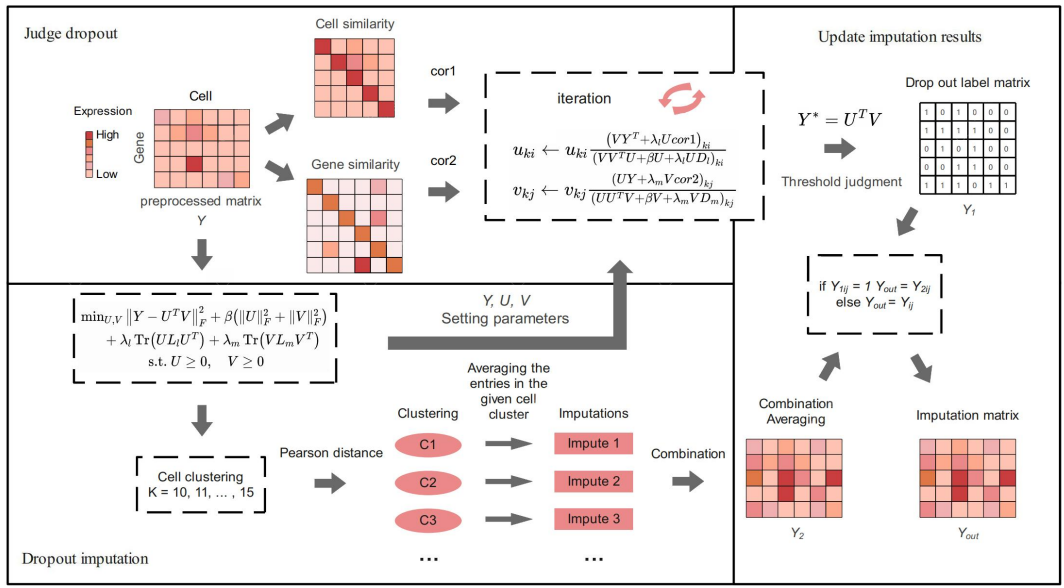

Figure 1. Overall model architecture of SFAG

Step 1: Technical zero value identification.

For the graph regularised non-negative matrix factorization objective function, this paper references the setting in GNMFLMI. The classical non-negative matrix factorization fails to discover the intrinsic geometric discriminant structure of the data space in Euclidean space. The classical non-negative matrix factorization is improved here.

Step 2: Average clustering results to impute the data.

Similar cells were first identified on the basis of clustering and imputation was performed by averaging the expression values of similar cells. To achieve robust estimates, multiple imputations are made using different cell clustering results and then multiple estimates are averaged to achieve the final imputation.

Step 3: Update imputation results.

The fill result obtained by the data smoothing fill algorithm is compared with the identified technology zero value and the fill result obtained in step 2 is updated to the location of the technology zero value.

## 2.2. Dataset

Two datasets from human(Darmanis[20]), and mouse(Deng[21]), were acquired to test the performance of SFAG.

## 3. MATERIALS AND METHODS

### 3.1. SFAG improved the performance of clustering method

Cluster analysis is one of the essential downstream analyses of scRNA-seq data. To show that SFAG can improve the clustering performance, We use ARI and NMI to measure the accuracy of cell clustering. We cluster the data imputed by the nine methods and the raw data, and the results are shown in Figure 2. As we can see from the plots, SFAG performs best on Darmanis, and is similar to the best performing approach on Deng.

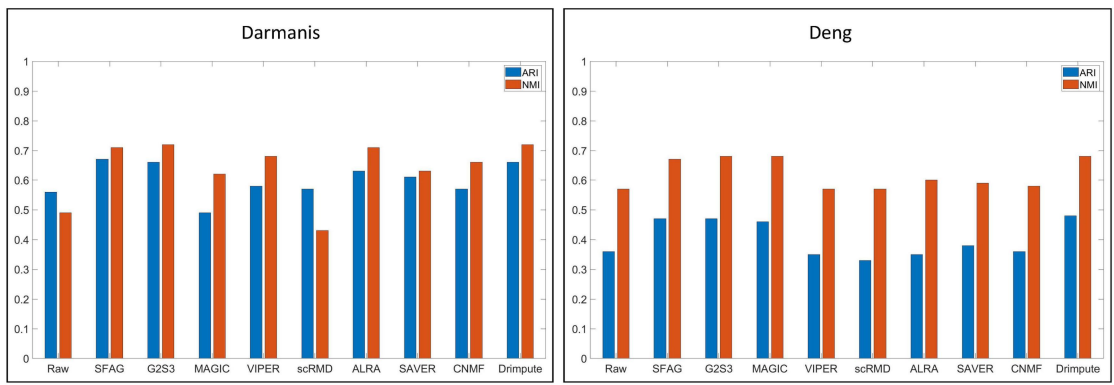

Figure 2. Comparison of ARI and NMI with different similarity measures

To more intuitively illustrate that SFAG can improve cluster analysis, we conducted a visual analysis on two datasets: Darmanis and Deng. We compare with eight methods and use UMAP to visualize the embeddwd representation in two-dimensional (2D), as shown in Figure 3. We can see from the plot that different types of cells can be well separated through the imputation of SFAG.

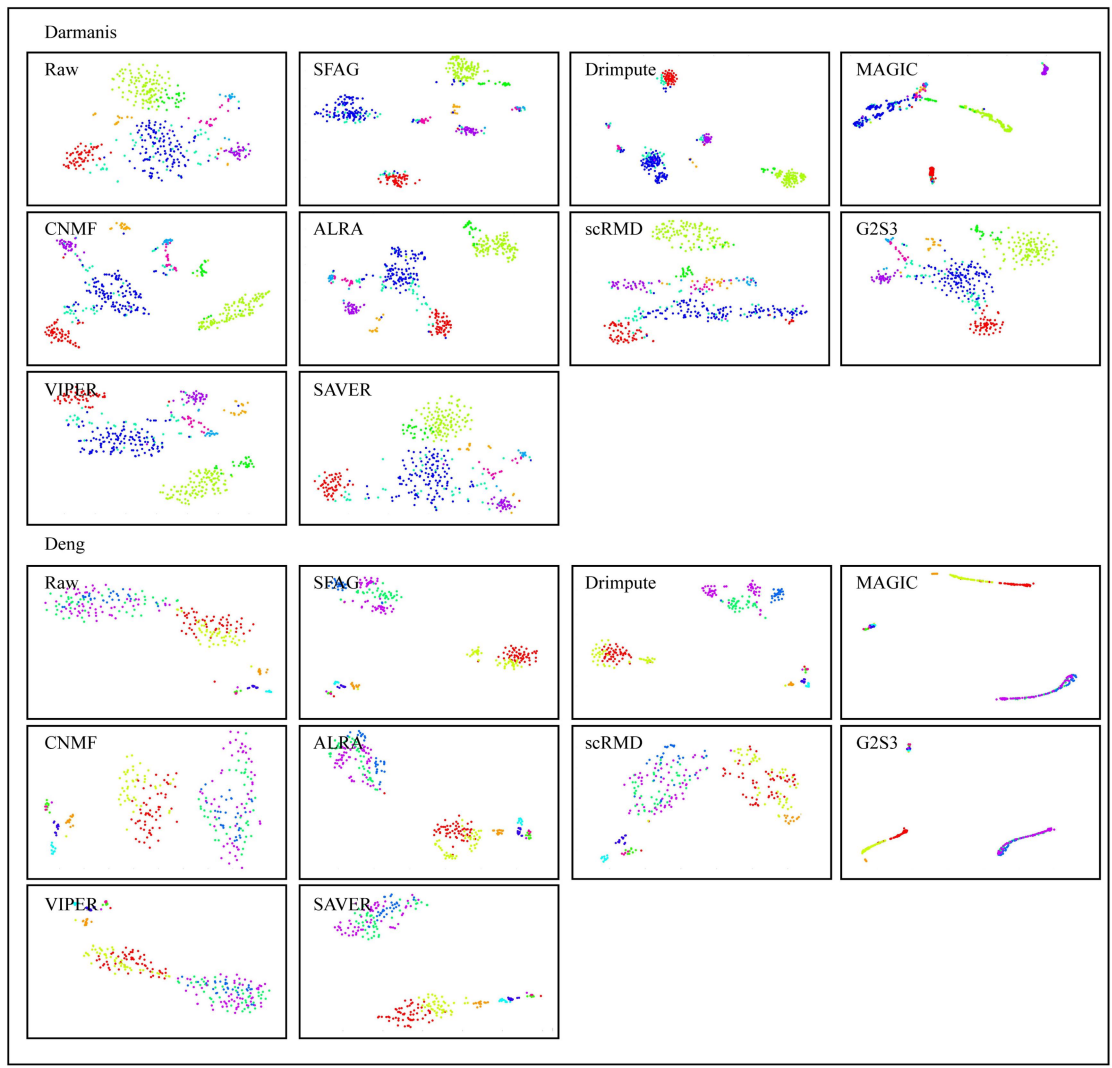

Figure 3. The umap visualization of cells from scRNA-seq datasets

## 3.2. Lineage reconstruction for cell trajectory

It is important and significant to study the cellular dynamic processes, which can be modeled computationally using pseudotime analysis. However, the performance of trajectory inference methods will be deteriorated due to dropout events. We evaluate the performance of different imputation methods by comparing the trajectory inference derived from different methods. and use KRCS and POS to measure the effect of trajectory analysis and the results are shown in Table 1. As we can see from the table, SFAG performs best on Darmanis.

Table 1. Heading and text fonts.

|  | Raw | SFAG | G2S3 | MAGIC | VIPER | scRMD | ALRA | SAVER | CNMF |
|---|---|---|---|---|---|---|---|---|---|
| KRSC | 0.388 | 0.639 | 0.576 | 0.387 | 0.287 | 0.499 | 0.477 | 0.271 | 0.301 |
| POS | 0.503 | 0.869 | 0.767 | 0.507 | 0.377 | 0.705 | 0.649 | 0.337 | 0.449 |

## 4. CONCLUSIONS

In this study, we propose a new imputation method by combining graph regularized nonnegative matrix factorization and data smoothing. We have carried out many experiments and found that it can detect false zeros more accurately than other methods and improve the clustering accuracy of ture data and reconstruct cell trajectory. However, SFAG still has some limitations. It is imputed without considering that the data may come from multiple batches. It may be the direction of our future research work.

## ACKNOWLEDGEMENTS

This work is supported in part by the National Natural Science Foundation of China (12126367, 12126305), the Hubei Provincial Natural Science Foundation of China under Grant 2015CFA010, the Fundamental Research Funds for the Central Universities, and the China University of Geosciences (Wuhan) under Grant CUGGC02.

## REFERENCES

[1] Tang, F & Barbacioru, C & Wang, Y, et al, (2008) mRNA-Seq whole-transcriptome analysis of a single cell, *Nature methods*, Vol. 6, No. 5, pp120-122.

[2] Shapiro, E & Biezuner, T & Linnarsson, S, (2013) Single-cell sequencing-based technologies will revolutionize whole-organism science, *Nature Reviews Genetics*, Vol. 14, No. 9, pp618-630.

[3] Patel, A, P & Tirosh, I & Trombetta, J, J & et al, (2014) Single-cell RNA-seq highlights intratumoral heterogeneity in primary glioblastoma, *Science*, Vol. 344, No. 6190, pp1396-1401.

[4] Zhou, F & Li, X & Wang, W & et al, (2016) Tracing haematopoietic stem cell formation at single-cell resolution, *Nature*, Vol. 533, No. 7604, pp487-492.

[5] Vallejos, C, A & Risso, D & Scialdone, A & et al, (2017) Normalizing single-cell RNA sequencing data: challenges and opportunities, *Nature methods*, Vol. 14, No. 6, pp565-571.

[6] Ziegenhain, C & Vieth, B & Parekh, S & et al, (2018) Quantitative single-cell transcriptomics, *Briefings in functional genomics*, Vol. 17, No. 4, pp220-232.

[7] Hou, W & Ji, Z & Ji, H & et al, (2020) A systematic evaluation of single-cell RNA-sequencing imputation methods, *Genome biology*, Vol. 21, pp1-30.

[8] Zhang, L & Zhang, S, (2018) Comparison of computational methods for imputing single-cell RNA-sequencing data, *IEEE/ACM transactions on computational biology and bioinformatics*, Vol. 17, No. 2, pp376-389.

[9] Van Dijk, D & Sharma, R & Nainys, J, et al, (2018) Recovering gene interactions from single-cell data using data diffusion, *Cell*, Vol. 174, No. 3, pp716-729.

[10] Li, W, V & Li, J, J, (2018) An accurate and robust imputation method scImpute for single-cell RNA-seq data, *Nature communications*, Vol. 9, No. 1, pp997.

[11] Huang, M & Wang, J & Torre, E & et al, (2018) SAVER: gene expression recovery for single-cell RNA sequencing, *Nature methods*, Vol. 15, No. 7, pp539-542.

[12] Xu, J & Cai, L & Liao, B & et al, (2020) CMF-Impute: an accurate imputation tool for single-cell RNA-seq data, *Bioinformatics*, Vol. 36, No. 10, pp3139-3147.

[13] Wang, M, N & You, Z, H & Li, L, P & et al, (2020) GNMFLMI: graph regularized nonnegative matrix factorization for predicting LncRNA-MiRNA interactions, *Ieee Access*, Vol. 8, pp37578-37588.

[14] Gong, W & Kwak, I, Y & Pota, P & et al, (2018) DrImpute: imputing dropout events in single cell RNA sequencing data, *BMC bioinformatics*, Vol. 19, pp1-10.

[15] Linderman, G, C & Zhao, J & Kluger, Y, (2018) Zero-preserving imputation of scRNA-seq data using low-rank approximation, *BioRxiv*.

[16] Chen, & Wu, C & Wu, L & et al, (2020) scRMD: imputation for single cell RNA-seq data via

robust matrix decomposition, *Bioinformatics*, Vol. 36, No. 10, pp3156-3161.

[17] Wu, W & Liu, Y & Dai, Q & et al, (2021) G2S3: A gene graph-based imputation method for single-cell RNA sequencing data, *PLOS Computational Biology*, Vol. 17, No. 5.

[18] Chen, M & Zhou, X, (2018) VIPER: variability-preserving imputation for accurate gene expression recovery in single-cell RNA sequencing studies, *Genome biology*, Vol. 19, No. 1, pp1-15.

[19] Vander, Maaten, L & Hinton, G, (2008) Visualizing data using t-SNE, *Journal of machine learning research*, Vol. 9, No. 11.

[20] Darmanis, S & Sloan, S, A & Zhang, Y & et al, (2015) A survey of human brain transcriptome diversity at the single cell level, *Proceedings of the National Academy of Sciences*, Vol. 112, No. 23, pp7285-7290.

[21] Deng, Q & Ramsköld, D & Reinius, B & et al, (2014) Single-cell RNA-seq reveals dynamic, random monoallelic gene expression in mammalian cells, *Science*, Vol. 343, No. 6167, pp193-196.